# Small amplitude ion-acoustic solitary waves in a four-component magneto-rotating plasma with a modified Cairns-Tsallis distribution

Hong Wang[1], Jiulin Du[1] [*], Ran Guo[2]

[1] *Department of Physics, School of Science, Tianjin University, Tianjin 300350, China*

[2] *Department of Physics, College of Science, Civil Aviation University of China, Tianjin 300300, China*

**Abstract** The small amplitude ion-acoustic solitary waves in the magneto-rotating plasma consisting of cold fluid ions, hot positrons, and the two-temperature electrons (cold and hot electrons) are investigated when the electrons obey a modified Cairns-Tsallis distribution. By using the reductive perturbation method, we derive the Korteweg–de Vries equation and the modified Korteweg–de Vries equation and obtain the small amplitude ion-acoustic solitary wave solutions. The dependences of solitary wave solutions on the nonextensive $q$-parameter, the nonthermal $\alpha$-parameter and the plasma physical quantities are analyzed numerically. We show the significant effects of the nonextensive $q$-parameter and the nonthermal $\alpha$-parameter etc. on the ion-acoustic solitary waves.

## 1. Introduction

Propagation of ion acoustic solitary waves (IASWs) is usually related to the plasma compression. For this reason, more solitary wave properties are obtained from the compressive/rarefactive waves in multi-component magnetized [1-4] or unmagnetized [5-7] plasmas by using the pseudo potential approach and the reductive perturbation method, i.e., Korteweg-deVries (KdV) soliton [8,9], in the linear [10] and nonlinear dynamics [11,12]. It has been revealed that the KdV soliton are formed due to a balance between dispersion and nonlinearity in the plasma. The dispersion effect is to make the wave packet diffuse, while the nonlinear effect is to make the wave packet converge. Propagation of the waves in the medium is unstable due to the dispersion. Only when a certain balance between the dispersion and the nonlinear convergence is reached, the solitary wave with stable waveform is generated [13].

In plasma, the dissipation is mainly caused by interparticle collisions, ion kinematic viscosity, and/or wave-particle interactions [9,14]. However, in some environments such as neutron stars, pulsars, quasars, black-hole magnetospheres and tokamak plasma, the plasmas are magnetized by a strong magnetic field. And the rapid rotation also increases the system's dissipation, where the Coriolis force may play a dominant role [4,15], so the propagation of IASWs in the plasma is usually studied in a non-inertial (rotating) frame. For example, Mushtaq and Shah studied the linear and nonlinear properties of IASWs in the magnetized rotating electron-positron-ion plasma [16], Khan and Iqbal studied solitons and shocks in the same plasma using the Homotopy perturbation method [17] and *tanh*-method, Farooq et al. investigated oblique drift solitary waves in a rotating electron-positron-ion plasma [18], Sahu et al. studied IASWs in a dense magneto-rotating quantum plasma [19]. Furthermore, Abbasi et al. analyzed the nonlinear ion-acoustic shock waves in the magneto-rotating relativistic plasmas with two-temperature superthermal electrons [20]. Jones et al. verified the linearity of ion-acoustic waves in the two-temperature plasma merging hot and cold electron groups [21]. Ghosh et al. studied the large Mach number ion acoustic rarefactive solitary waves by using Sagdeev pseudopotential in the plasma with two-temperature electrons [22]. Sabry developed the theory of IASWs and double-layer in the electron-positron-ion plasma with two-temperature electrons [5].

---

[*] Corresponding author, E-mail: jldu@tju.edu.cn

In the above works, it was generally assumed the two-temperature electrons to be a Maxwellian distribution. The Maxwell-distributed electrons are monoenergetic electron beam, which may reach a state of thermal equilibrium due to energy loss from collisions of plasma components. The collisional relaxation time depends on the energy balance process and the impact frequency of electrons [23]. However, as more and more empirical data became available, people realized that in actual plasma systems, the velocity distribution of particles deviates significantly from the Maxwellian distribution. Therefore, when using the plasma models based on a Maxwellian distribution to explain or predict the waves and instabilities in complex plasmas, it does not give a good quantitative fit with the observed results. In addition, in the excitation of waves, the strong interactions can also keep the trapped electrons away from the thermal equilibrium state [24]. In fact, the non-Maxwellian velocity distributions are very common both in laboratory plasmas and astrophysical and space plasmas, such as the solar flares, the magneto tail, near plasma shock waves [25], the Earth's plasma sheets [26] and the solar wind [27] etc., which lead to recent many interesting works in the multi-component complex plasmas with power-law /non-Maxwellian velocity distributions. For example, the dust-acoustic waves and stability in the permeating plasma obeying the power-law *q*-distribution [28], the Zakharov-Kuznetsov equation in a magnetized plasma containing kappa-distributed hot and cold electrons [29], the super-periodicity, chaos and coexisting orbits of ion-acoustic waves in the multi-component plasma consisting of fluid ions, nonextensive cold and hot electrons [30].

In order to simulate the velocity distribution of the superthermal particles observed by the Freja satellite [31] and Viking spacecraft [32], Cairns distribution, a non-Maxwellian distribution with an enhanced high-energy tail, was introduced to characterize the deviation from the Maxwellian velocity distribution [33]. The Cairns distribution was applied to investigate oblique propagating of IASWs in a magnetized plasma consisting of warm adiabatic ions and nonthermal electrons, where the reductive perturbation method was used to determine the solitary wave solution with small amplitude [34]. And it was also applied to study electron acoustic solitary waves in the plasma with nonthermal hot electrons, positrons and cold dynamical electrons [35].

In order to fit the observations of astrophysical and space plasmas more accurately, a hybrid Cairns-Tsallis velocity distribution was proposed from nonextensive statistics [36, 37], which provides the enhanced parameter flexibility for modeling the nonthermal plasmas with two parameters. There have been many works on the complex plasmas with the Cairns-Tsallis distribution [18,38,39], such as the modulational instability of electron-acoustic waves [7], the nonlinear dust acoustic waves [37], the effect of polarization force on oppositely charged dust grains in a magnetized plasma [38]. Most recently, based on the basic probability independence postulate in nonextensive statistics, a modified Cairns-Tsallis distribution was applied to the multi-component plasma with a potential [40].

Since most astrophysical and space plasmas are multi-component, and they are often magnetized, for more realistic images it is necessary to study ion-acoustic waves in a non-inertial frame. In order to further study the effects of magnetic field and rotation on IASWs, in this work, we use the reductive perturbative method to study the nonlinear IASWs in the four-component plasma consisting of ions, two-temperature electrons and positrons, where the two-temperature electrons follow a modified Cairns-Tsallis distribution. Here we focus on the small amplitude IASWs. In Section 2, we give the basic plasma model and hydrodynamic equations. In Section 3 and Section 4, we use the reductive perturbative method to study the KdV equation, the modified KdV equation and the corresponding small amplitude soliton solutions, where the numerical analyses are made. In Section 5, we give the conclusion.

## 2. The plasma model and governing equations

We consider the three-dimensional collisionless magneto-rotating plasma, composed of cold fluid ions, hot positrons and the cold and hot electrons (two-temperature electrons) which obey the modified Cairns-Tsallis distribution. We use $n_i$, $n_p$, $n_c$ and $n_h$ to denote the particle number density of the ions, the positrons, the cold electrons and the hot electrons, respectively, and use $T_i$, $T_p$, $T_c$ and $T_h$ to denote their corresponding temperatures. The external magnetic field direction is taken along the z-axis, i.e., $\mathbf{B} = B_0 \hat{z}$, where $\hat{z}$ is the unit vector along the z-axis. The plasma is assumed to be rotating slowly around the z-axis with an angular frequency $\Omega$ ($\Omega < 1$), i.e., $\mathbf{\Omega} = \Omega_0 \hat{z}$, such that the quadratic and higher terms of $\Omega$ can be ignored in the rotational plasma, and the centrifugal force $\sim \mathbf{\Omega} \times (\mathbf{\Omega} \times \mathbf{r})$ effect may be neglected. In the ion-acoustic waves, inertias of the electron and positron are neglected, and the electron and positron fluids can be separately in equilibrium with an electrostatic potential $\varphi$. For such a plasma model, the nonlinear dynamics of low-frequency ion-acoustic waves are governed by the following equations [1,2,5,41],

$$\frac{\partial n_i}{\partial t} + \nabla \cdot (n_i \mathbf{u}) = 0, \tag{1}$$

$$\frac{\partial \mathbf{u}}{\partial t} + (\mathbf{u} \cdot \nabla)\mathbf{u} = -\nabla \Psi + \varpi_{ci} \mathbf{u} \times \mathbf{z} + 2\mathbf{u} \times \mathbf{\Omega} - \frac{\sigma}{n_i} \nabla p, \tag{2}$$

$$\frac{\partial p}{\partial t} + (\mathbf{u} \cdot \nabla) p + \gamma p \nabla \cdot \mathbf{u} = 0, \tag{3}$$

$$\nabla^2 \Psi = n_h + n_c - \eta n_p - (1-\eta) n_i, \tag{4}$$

where $\mathbf{u} = (u_x, u_y, u_z)$ is the fluid velocity of ions normalized by the effective ion-acoustic speed $c_s = (kT_{eff}/m_i)^{1/2}$ with Boltzmann constant $k$, the effective temperature $T_{eff}$ of two-temperature electrons and the ion mass $m_i$. If $n_{e0}$, $n_{c0}$, $n_{h0}$, $n_{p0}$ and $n_{i0}$ are the equilibrium number density of the total electrons, the cold electron, the hot electrons, the positrons and the ions, respectively, then the densities $n_c$, $n_h$ and $n_p$ can be normalized by $n_{e0}$; $\eta = n_{p0}/n_{e0}$ is the equilibrium density ratio of the positrons to the total electrons; the fluid pressure $p$ is normalized by the equilibrium pressure [42], $p_0 = n_{i0} k T_i$; the space variables ($x$, $y$, $z$) are normalized by the effective Debye length, $\lambda_{De} = (kT_{eff}/4\pi n_{e0} e^2)^{1/2}$; the time $t$ is normalized by the ion plasma period, $\Omega_p^{-1} = \sqrt{m_i/4\pi e^2 n_{e0}}$; the ion gyrofrequency $\varpi_{ci} = eB_0/(m_i c)$ is normalized by $\Omega_p$; $\Psi = -e\varphi/kT_{eff}$ is the normalized electrostatic potential; $e$ is the electronic charge and $c$ is the speed of light; $\sigma = T_i/T_{eff}$ is temperature ratio of the ions to the two-temperature electrons; and $\gamma$ is $\gamma = C_p^i/C_v^i$, where $C_p^i$ (or $C_v^i$) is the specific heat of the ion at constant pressure (or volume) [1-3]. In the above statements, the effective temperature of the cold and hot electrons (two-temperature electrons) in the plasma is $T_{eff} = T_c/(\mu_c + \mu_h \beta)$ if $\beta = T_c/T_h$ is the temperature ratio of the cold electrons to the hot electrons, $\mu_c = n_{c0}/n_{e0}$ and $\mu_h = n_{h0}/n_{e0}$ are the density ratios of the cold and hot electrons, respectively, to the total electrons at $\phi = 0$ (so we have $\mu_c + \mu_h = 1$).

For the present four-component magneto-rotating plasma with the two-temperature electrons following the Cairns-Tasllis distribution, when we consider the Tsallis $q$-distribution with a potential $\varphi(r)$ in nonextensive statistics, the modified Cairns-Tasllis distribution of the $j$th component ($j = c$, $h$ is the cold and hot electrons, respectively) should be written [40, 43-44] as

$$f(r_j, v_j) = C_j(q, \alpha) \left[ 1 + \alpha \left( \frac{m_j v_j^2 + 2e\varphi(r_j)}{kT_j} \right)^2 \right] \left[ 1 - (q-1)\frac{m_j v_j^2}{2kT_j} \right]^{\frac{1}{q-1}} \left[ 1 - (q-1)\frac{e\varphi(r_j)}{kT_j} \right]^{\frac{1}{q-1}} \tag{5}$$

where the normalization coefficient is that

$$C_j(q,\alpha) = \frac{n_{j0}}{(2\pi)^{3/2} v_{te}^3} \begin{cases} \dfrac{(1-q)^{5/2} \Gamma\left(\dfrac{1}{1-q}\right)}{\Gamma\left(\dfrac{1}{1-q}-\dfrac{5}{2}\right)\left[3\alpha+(1-q)^2\left(\dfrac{1}{1-q}-\dfrac{3}{2}\right)\left(\dfrac{1}{1-q}-\dfrac{5}{2}\right)\right]}, & \dfrac{3}{5} < q < 1, \\[2em] \dfrac{(q-1)^{5/2}\left(\dfrac{1}{q-1}+\dfrac{5}{2}\right)\left(\dfrac{1}{q-1}+\dfrac{3}{2}\right)\Gamma\left(\dfrac{1}{q-1}+\dfrac{3}{2}\right)}{\left[(q-1)^2\left(\dfrac{1}{q-1}+\dfrac{3}{2}\right)\left(\dfrac{1}{q-1}+\dfrac{5}{2}\right)+3\alpha\right]\Gamma\left(\dfrac{1}{q-1}+1\right)}, & q > 1. \end{cases}$$

In (5), $v_{te} = (kT_e/m_e)^{1/2}$ is the thermal velocity of electrons, $m_j$, $v_j$ and $T_j$ are the mass, velocity and temperature, respectively. The nonthermal parameter $\alpha > 0$ represents the number of nonthermal electrons [33]. The nonextensive parameter is $q > 0$ and $q \neq 1$, who's deviation from unity describes the nonextensive degree of the system. In a nonequilibrium rotating plasma, $q$ can be determined by the equation [45, 46],

$$k\nabla T_j = (q-1)\left[e(\nabla\varphi - c^{-1}\mathbf{u}\times\mathbf{B}) + m_j(\Omega^2\mathbf{R} + 2\mathbf{u}\times\mathbf{\Omega})\right], \qquad (6)$$

where $\mathbf{R}$ is the vertical distance between the particle and the rotation axis, and its orientation is outward and perpendicular to the axis. It is worth noting that for $q > 1$, there is a thermal cutoff on the maximum value allowed for the velocity in the distribution (5), namely, $v_{max} = \sqrt{2kT_j/m_j(q-1)}$, while for $q < 1$, there is no limit for the velocity.

The hybrid Cairns-Tsallis distribution has obvious advantage since Cairns distribution could be modified in the low energy as well as in the high energy parts by changing the nonthermal parameter $\alpha$, whereas Tsallis distributions also has the similar characteristics by changing the nonextensive parameter $q$. In the limit $q \to 1$, the distribution (5) returns to the Cairns distribution [33]. When we take $\alpha = 0$, (5) becomes the Tsallis $q$-distribution, and when we take $q \to 1$ and $\alpha = 0$, it recovers the Maxwellian distribution. And if the potential $\varphi = 0$, (5) becomes the same as the original Cairns-Tsallis velocity distribution. In Figure 1(a), we plotted the distribution function (5) as a function of U= $v/v_{te}$ for five different values of the nonextensive $q$-parameter and one fixed value of the nonthermal $\alpha$-parameter, where $q=1$ is corresponding to a case of the nonthermal Cairns distribution. We show that, as $q$ increases, the shoulders of the distribution function become narrower and the probability of high energy states becomes smaller. In Figure 1(b), we plotted the distribution function (5) as a function of U= $v/v_{te}$ for four different values of the nonthermal $\alpha$–parameter and one fixed value of the nonextensive $q$-parameter, where $\alpha = 0$ is corresponding to a case of the *q*-distribution in nonextensive statistics. We show that, with the increase of *α*, the shoulders of the distribution function become more prominent, the probability of high energy states is greater, and the probability of low energy states is less.

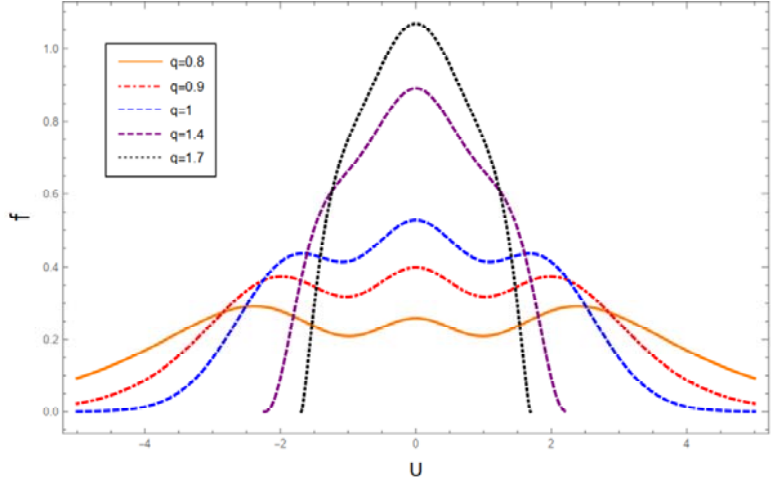

**Figure 1(a).** The modified CT distribution (5) with *α* = 0.3 for five different *q*.

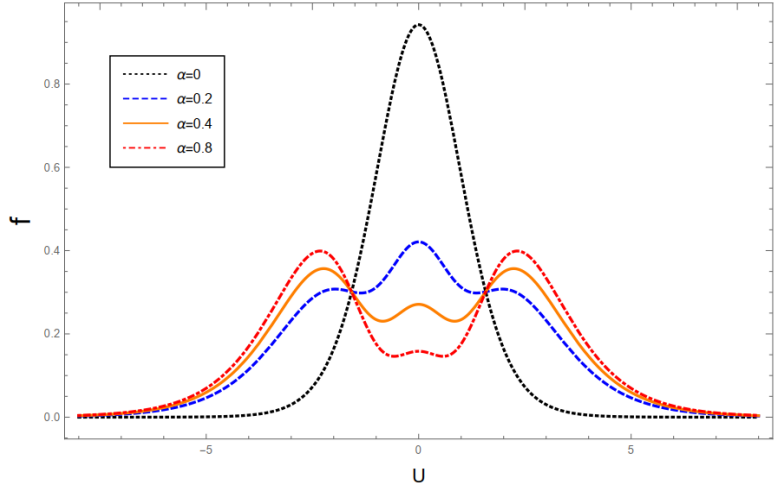

**Figure 1(b).** The modified CT distribution (5) with *q* = 0.85 for four different *α*.

By using the unnormalized electrostatic potential *φ* and then integrating the distribution function (5) for the velocity space, the electron number density is obtained [40] as

$$n_j(\varphi) = n_{j0}\left[1-(q-1)\frac{e\varphi}{kT_j}\right]^{\frac{1}{q-1}}\left[1-A\frac{e\varphi}{kT_j}+B\left(\frac{e\varphi}{kT_j}\right)^2\right], \quad (7)$$

where the abbreviations *A* and *B* are

$$A = -\frac{8\alpha(5q-3)}{(3q-1)(5q-3)+12\alpha} \quad \text{and} \quad B = \frac{4\alpha(3q-1)(5q-3)}{(3q-1)(5q-3)+12\alpha}.$$

In Eq. (7), if the electrostatic potential $\varphi$ is replaced by the normalized $\Psi$, the number density of the cold and hot electrons can be written, respectively, as

$$n_c(\Psi) = \mu_c \left[1 + \frac{(q-1)\Psi}{\mu_c + \mu_h \beta}\right]^{\frac{1}{q-1}} \left[1 + \frac{A\Psi}{\mu_c + \mu_h \beta} + B\left(\frac{\Psi}{\mu_c + \mu_h \beta}\right)^2\right], \tag{8}$$

$$n_h(\Psi) = \mu_h \left[1 + \frac{(q-1)\beta\Psi}{\mu_c + \mu_h \beta}\right]^{\frac{1}{q-1}} \left[1 + \frac{A\beta\Psi}{\mu_c + \mu_h \beta} + B\left(\frac{\beta\Psi}{\mu_c + \mu_h \beta}\right)^2\right], \tag{9}$$

where we denote $\mu_c = n_{c0}/n_{e0}$, $\mu_h = n_{h0}/n_{e0}$, $\beta = T_c/T_h$, $T_{eff} = T_c/(\mu_c + \mu_h \beta)$ and $\Psi = -e\varphi/k_B T_{eff}$. Obviously, $\mu_c + \mu_h = 1$. Further, if $\nu = T_{eff}/T_p$ is a ratio of the effective temperature to the temperature of positrons, then the number density of positrons can be expressed as

$$n_p = \exp(-\nu\Psi). \tag{10}$$

In the present model, we focus on the small amplitude IASWs for $\Psi \ll 1$, so we can expand the densities $n_c$, $n_h$ and $n_p$ into a power series of $\Psi$, namely,

$$n_c \approx \mu_c\left(1 + C_1\Psi + C_2\Psi^2 + C_3\Psi^3 + ...\right), \tag{11}$$

$$n_h \approx \mu_h\left(1 + C_1\beta\Psi + C_2\beta^2\Psi^2 + C_3\beta^3\Psi^3 + ...\right), \tag{12}$$

$$n_p \approx 1 - \nu\Psi + \frac{\nu^2}{2}\Psi^2 - \frac{\nu^3}{6}\Psi^3 + ..., \tag{13}$$

where the coefficients $C_1$, $C_2$ and $C_3$ are

$$C_1 = \frac{1+A}{\mu_c + \mu_h\beta}, \quad C_2 = \frac{2+2A+2B-q}{2(\mu_c + \mu_h\beta)^2} \quad \text{and} \quad C_3 = \frac{6+6A+6B-7q-3Aq+2q^2}{6(\mu_c + \mu_h\beta)^3}. \tag{14}$$

Therefore, Poisson's equation (4) can be written by the following form,

$$\nabla^2\phi = \mu_h\left(1 + C_1\beta\Psi + C_2\beta^2\Psi^2 + C_3\beta^3\Psi^3\right) + \mu_c\left(1 + C_1\Psi + C_2\Psi^2 + C_3\Psi^3\right)$$
$$-\eta\left(1 - \nu\Psi + \frac{\nu^2}{2}\Psi^2 - \frac{\nu^3}{6}\Psi^3\right) - (1-\eta)n_i. \tag{15}$$

For convenience to calculate, we further write the governing Eqs. (1)-(4) as

$$\frac{\partial n_i}{\partial t} + \frac{\partial(n_i u_x)}{\partial x} + \frac{\partial(n_i u_y)}{\partial y} + \frac{\partial(n_i u_z)}{\partial z} = 0, \tag{16}$$

$$\frac{\partial u_x}{\partial t} + u_x\frac{\partial u_x}{\partial x} + u_y\frac{\partial u_x}{\partial y} + u_z\frac{\partial u_x}{\partial z} = -\frac{\partial \Psi}{\partial x} + \Omega_{eff}u_y - \frac{\sigma}{n_i}\frac{\partial p}{\partial x}, \tag{17}$$

$$\frac{\partial u_y}{\partial t} + u_x\frac{\partial u_y}{\partial x} + u_y\frac{\partial u_y}{\partial y} + u_z\frac{\partial u_y}{\partial z} = -\frac{\partial \Psi}{\partial y} - \Omega_{eff}u_x - \frac{\sigma}{n_i}\frac{\partial p}{\partial y}, \tag{18}$$

$$\frac{\partial u_z}{\partial t} + u_x\frac{\partial u_z}{\partial x} + u_y\frac{\partial u_z}{\partial y} + u_z\frac{\partial u_z}{\partial z} = -\frac{\partial \Psi}{\partial z} - \frac{\sigma}{n_i}\frac{\partial p}{\partial z}, \tag{19}$$

$$\frac{\partial p}{\partial t}+\left(u_x\frac{\partial p}{\partial x}+u_y\frac{\partial p}{\partial y}+u_z\frac{\partial p}{\partial z}\right)+\gamma p\left(\frac{\partial u_x}{\partial x}+\frac{\partial u_y}{\partial y}+\frac{\partial u_z}{\partial z}\right)=0, \quad (20)$$

$$\left(\frac{\partial^2}{\partial x^2}+\frac{\partial^2}{\partial y^2}+\frac{\partial^2}{\partial z^2}\right)\Psi = \mu_h\left(1+C_1\beta\Psi+C_2\beta^2\Psi^2+C_3\beta^3\Psi^3\right)+\mu_c\left(1+C_1\Psi+C_2\Psi^2+C_3\Psi^3\right)$$
$$-\eta\left(1-\nu\Psi+\frac{\nu^2}{2}\Psi^2-\frac{\nu^3}{6}\Psi^3\right)-(1-\eta)n_i, \quad (21)$$

where $\Omega_{eff}=\varpi_{ci}+2\Omega_0$.

## 3. The KdV equation and the IASWs

### 3.1 *The equation*

The standard reduction perturbation method (RPM) is usually suitable for the study of small but finite amplitude solitary waves [47]. Therefore, we apply the RPM to Eqs. (16)-(21) to derive the nonlinear KdV equation of small amplitude IASWs in the four-component magneto-rotating plasma with two-temperature electrons. According to the RPM, the independent variables ($x$, $y$, $z$, $t$) are stretched to ($\xi$, $\tau$) by

$$\xi=\varepsilon^{1/2}\left(l_x x+l_y y+l_z z-\lambda t\right), \quad \tau=\varepsilon^{3/2}t, \quad (22)$$

where $\varepsilon$ is a small parameter ($0<\varepsilon\ll 1$) characterizing strength of the nonlinearity, $\lambda$ is phase speed of the waves, $l_x$, $l_y$ and $l_z$ are the directional cosines of the wave vector **k** along the $x$, $y$ and $z$ axes, respectively, thus we have that $l_x^2+l_y^2+l_z^2=1$. Now we can write the physical quantities as the power series about the small parameter $\varepsilon$, namely,

$$n_i=1+\varepsilon n_i^{(1)}+\varepsilon^2 n_i^{(2)}+\mathrm{L},$$
$$u_x=\varepsilon^{3/2}u_x^{(1)}+\varepsilon^2 u_x^{(2)}+\mathrm{L},$$
$$u_y=\varepsilon^{3/2}u_y^{(1)}+\varepsilon^2 u_y^{(2)}+\mathrm{L},$$
$$u_z=\varepsilon u_z^{(1)}+\varepsilon^2 u_z^{(2)}+\mathrm{L},$$
$$p=1+\varepsilon p^{(1)}+\varepsilon^2 p^{(2)}+\mathrm{L},$$
$$\Psi=\varepsilon\Psi^{(1)}+\varepsilon^2\Psi^{(2)}+\mathrm{L}. \quad (23)$$

where $u_x^{(1)}$ and $u_y^{(1)}$ are smaller due to the drift $\mathbf{E}\times\mathbf{B_0}$ in a magnetized plasma [1,2,16]. Using Eq. (22), the derivatives of position $r=(x, y, z)$ and time $t$ become that

$$\frac{\partial}{\partial r}=\varepsilon^{1/2}l_r\frac{\partial}{\partial\xi}, \quad \frac{\partial^2}{\partial r^2}=\varepsilon l_r^2\frac{\partial^2}{\partial\xi^2}, \quad \text{and} \quad \frac{\partial}{\partial t}=-\lambda\varepsilon^{1/2}\frac{\partial}{\partial\xi}+\varepsilon^{3/2}\frac{\partial}{\partial\tau}, \quad (24)$$

and then, Eqs. (16)-(21) become that

$$-\lambda\varepsilon^{1/2}\frac{\partial n_i}{\partial\xi}+\varepsilon^{3/2}\frac{\partial n_i}{\partial\tau}+\varepsilon^{1/2}l_x\frac{\partial(n_i u_x)}{\partial\xi}+\varepsilon^{1/2}l_y\frac{\partial(n_i u_y)}{\partial\xi}+\varepsilon^{1/2}l_z\frac{\partial(n_i u_z)}{\partial\xi}=0, \quad (25)$$

$$\varepsilon^{3/2}\frac{\partial u_x}{\partial \tau}+\varepsilon^{1/2}\left(-\lambda+u_x l_x+u_y l_y+u_z l_z\right)\frac{\partial u_x}{\partial \xi}+\varepsilon^{1/2}l_x\frac{\partial \Psi}{\partial \xi}-\Omega_{eff}u_y+\varepsilon^{1/2}\frac{\sigma}{n_i}l_x\frac{\partial p}{\partial \xi}=0, \qquad (26)$$

$$\varepsilon^{3/2}\frac{\partial u_y}{\partial \tau}+\varepsilon^{1/2}\left(-\lambda+u_x l_x+u_y l_y+u_z l_z\right)\frac{\partial u_y}{\partial \xi}+\varepsilon^{1/2}l_y\frac{\partial \Psi}{\partial \xi}+\Omega_{eff}u_x+\varepsilon^{1/2}\frac{\sigma}{n_i}l_y\frac{\partial p}{\partial \xi}=0, \qquad (27)$$

$$\varepsilon^{3/2}\frac{\partial u_z}{\partial \tau}+\varepsilon^{1/2}\left(-\lambda+u_x l_x+u_y l_y+u_z l_z\right)\frac{\partial u_z}{\partial \xi}+\varepsilon^{1/2}l_z\frac{\partial \Psi}{\partial \xi}+\varepsilon^{1/2}\frac{\sigma}{n_i}l_z\frac{\partial p}{\partial \xi}=0, \qquad (28)$$

$$\varepsilon^{3/2}\frac{\partial p}{\partial \tau}+\varepsilon^{1/2}\left(-\lambda+u_x l_x+u_y l_y+u_z l_z\right)\frac{\partial p}{\partial \xi}+\varepsilon^{1/2}\gamma p\left(l_x\frac{\partial u_x}{\partial \xi}+l_y\frac{\partial u_y}{\partial \xi}+l_z\frac{\partial u_z}{\partial \xi}\right)=0, \qquad (29)$$

$$\varepsilon\frac{\partial^2 \Psi}{\partial \xi^2}=-\eta+\aleph\Psi+\Re\Psi^2+\Xi\Psi^3-(1-\eta)n_i, \qquad (30)$$

where we have denoted that

$$\aleph=\mu_h C_1\beta+\mu_c C_1+\eta\nu, \quad \Re=\mu_h C_2\beta^2+\mu_c C_2-\frac{\eta\nu^2}{2}, \quad \text{and} \quad \Xi=\mu_h C_3\beta^3+\mu_c C_3+\frac{\eta\nu^3}{6}.$$

After substituting Eq. (23) into Eqs. (25)-(30), for the first-terms of the small parameter $\varepsilon$, we get

$$n_i^{(1)}=\frac{\aleph}{1-\eta}\Psi^{(1)}, \quad n_i^{(1)}=\frac{l_z}{\lambda}u_z^{(1)}, \quad p^{(1)}=\frac{\gamma l_z}{\lambda}u_z^{(1)}$$

$$u_y^{(1)}=\frac{1}{\Omega_{eff}}\left(l_x\frac{\partial \Psi^{(1)}}{\partial \xi}+\sigma l_x\frac{\partial p^{(1)}}{\partial \xi}\right), \quad u_x^{(1)}=\frac{1}{\Omega_{eff}}\left(-\sigma l_y\frac{\partial p^{(1)}}{\partial \xi}-l_y\frac{\partial \Psi^{(1)}}{\partial \xi}\right),$$

$$u_z^{(1)}=\frac{l_z}{\lambda}\Psi^{(1)}+\frac{\sigma l_z}{\lambda}p^{(1)}, \qquad (31)$$

where we have used the boundary conditions [48], i.e., the first-order perturbations $\left(n_i^{(1)},u_j^{(1)},\Psi^{(1)}\right)$ tend to zero at $\xi\rightarrow\pm\infty$. Moreover, we find that the phase speed can be expressed as

$$\lambda=l_z\sqrt{\frac{1-\eta}{\aleph}+\sigma\gamma}. \qquad (32)$$

Now we express these first-order quantities as a function of $\Psi^{(1)}$, namely,

$$n_i^{(1)}=\frac{l_z^2}{\lambda^2-\sigma\gamma l_z^2}\Psi^{(1)}, \quad p^{(1)}=\frac{\gamma l_z^2}{\lambda^2-\sigma\gamma l_z^2}\Psi^{(1)}, \quad u_z^{(1)}=\frac{\lambda l_z}{\lambda^2-\sigma\gamma l_z^2}\Psi^{(1)},$$

$$u_x^{(1)}=-\frac{l_y}{\Omega_{eff}}\left(1+\frac{\sigma\gamma l_z^2}{\lambda^2-\sigma\gamma l_z^2}\right)\frac{\partial \Psi^{(1)}}{\partial \xi}, \quad u_y^{(1)}=\frac{l_x}{\Omega_{eff}}\left(1+\frac{\sigma\gamma l_z^2}{\lambda^2-\sigma\gamma l_z^2}\right)\frac{\partial \Psi^{(1)}}{\partial \xi}. \qquad (33)$$

For the second-order terms of $\varepsilon$, we get

$$n_i^{(2)}=\frac{\aleph}{1-\eta}\Psi^{(2)}+\frac{\Re}{1-\eta}\left[\Psi^{(1)}\right]^2-\frac{1}{1-\eta}\frac{\partial^2 \Psi^{(1)}}{\partial \xi^2},$$

$$-\lambda\frac{\partial n_i^{(2)}}{\partial \xi}+\frac{\partial n_i^{(1)}}{\partial \tau}+l_x\frac{\partial u_x^{(2)}}{\partial \xi}+l_y\frac{\partial u_y^{(2)}}{\partial \xi}+l_z\frac{\partial u_z^{(2)}}{\partial \xi}+l_z\frac{\partial\left(n_i^{(1)}u_z^{(1)}\right)}{\partial \xi}=0,$$

$$u_y^{(2)}=-\frac{\lambda}{\Omega_{eff}}\frac{\partial u_x^{(1)}}{\partial \xi}, \quad u_x^{(2)}=\frac{\lambda}{\Omega_{eff}}\frac{\partial u_y^{(1)}}{\partial \xi},$$

$$\frac{\partial u_z^{(1)}}{\partial \tau} - \lambda \frac{\partial u_z^{(2)}}{\partial \xi} + l_z u_z^{(1)} \frac{\partial u_z^{(1)}}{\partial \xi} - \lambda n_i^{(1)} \frac{\partial u_z^{(1)}}{\partial \xi} + l_z \frac{\partial \Psi^{(2)}}{\partial \xi} + l_z n_i^{(1)} \frac{\partial \Psi^{(1)}}{\partial \xi} + \sigma l_z \frac{\partial p^{(2)}}{\partial \xi} = 0,$$

$$\frac{\partial p^{(1)}}{\partial \tau} + l_z u_z^{(1)} \frac{\partial p^{(1)}}{\partial \xi} - \lambda \frac{\partial p^{(2)}}{\partial \xi} + \gamma l_x \frac{\partial u_x^{(2)}}{\partial \xi} + \gamma l_y \frac{\partial u_y^{(2)}}{\partial \xi} + \gamma l_z \frac{\partial u_z^{(2)}}{\partial \xi} + \gamma l_z p^{(1)} \frac{\partial u_z^{(1)}}{\partial \xi} = 0. \qquad (34)$$

From Eqs. (31)-(34), we can derive the KdV equation for ion-acoustic waves in the plasma, i.e.,

$$\frac{\partial \Psi^{(1)}}{\partial \tau} + P_1 \Psi^{(1)} \frac{\partial \Psi^{(1)}}{\partial \xi} + P_2 \frac{\partial^3 \Psi^{(1)}}{\partial \xi^3} = 0, \qquad (35)$$

where the nonlinear coefficient $P_1$ and the dispersion coefficient $P_2$ are given by

$$P_1 = -\frac{\Re(\lambda^2 - \sigma\gamma l_z^2)^2}{\lambda l_z^2 (1-\eta)} + \frac{3l_z^2}{2\lambda} + \frac{\sigma\gamma(1+\gamma)l_z^4}{2\lambda(\lambda^2 - \sigma\gamma l_z^2)}, \qquad (36)$$

$$P_2 = \frac{(\lambda^2 - \sigma\gamma l_z^2)^2}{2\lambda l_z^2 (1-\eta)} + \frac{\lambda^3 (1-l_z^2)}{2l_z^2 \Omega_{eff}^2}. \qquad (37)$$

3.2 *The IASWs*

If we let $\Psi^{(1)} = \Psi$ and $\chi = \xi - U_0 \tau$, where $U_0$ is the solitary wave speed in a moving frame, then, the KdV equation (35) becomes the ordinary differential equation,

$$-U_0 \frac{d\Psi}{d\chi} + P_1 \Psi \frac{d\Psi}{d\chi} + P_2 \frac{d^3\Psi}{d\chi^3} = 0. \qquad (38)$$

By integrating (38) over $\chi$ and then multiplying by $d\Psi/d\chi$ to integrate $\chi$ again, we get

$$-\frac{1}{2}U_0 \Psi^2 + \frac{1}{6} P_1 \Psi^3 + \frac{1}{2} P_2 \left(\frac{d\Psi}{d\chi}\right)^2 = c_1 \Psi + c_2, \qquad (39)$$

where $c_1$ and $c_2$ are the integration constants. The solitary wave solution is a locally stable that decays to zero at infinity, thus there are boundary conditions at $\chi \to \pm\infty$,

$$\Psi \to 0, \quad d\Psi/d\chi \to 0, \quad d^2\Psi/d\chi^2 \to 0. \qquad (40)$$

Using boundary condition (40), we determine that $c_1 = c_2 = 0$, so Eq. (39) becomes

$$\left(\frac{d\Psi}{d\chi}\right)^2 = \Psi^2 \left(\frac{U_0}{P_2} - \frac{P_1}{3P_2}\Psi\right). \qquad (41)$$

Integrating Eq. (41) and using the boundary condition (40) again, we get the solution,

$$\chi = \mp \frac{2\operatorname{ArcTanh}\left[\sqrt{1 - \frac{P_1 \Psi}{3U_0}}\right]\sqrt{3U_0 - P_1 \Psi}}{\sqrt{U_0 \left(\frac{3U_0 - P_1 \Psi}{P_2}\right)}} = \mp \sqrt{\frac{4P_2}{U_0}} \operatorname{ArcTanh}\left[\sqrt{1 - \frac{P_1 \Psi}{3U_0}}\right]. \qquad (42)$$

Therefore, we find the IASWs solution of the KdV equation (35),

$$\Psi = \frac{3U_0}{P_1}\left[1-\text{Tanh}^2\left(\frac{\chi}{2}\sqrt{\frac{U_0}{P_2}}\right)\right] = \frac{3U_0}{P_1}\text{Sech}^2\left(\frac{\chi}{2}\sqrt{\frac{U_0}{P_2}}\right). \tag{43}$$

As compared with the standard form of solitary wave solutions [49,50], i.e., $\Psi = \Psi_m \text{sech}^2[\chi/\Delta]$, we find the amplitude and width of the IASWs, respectively, $\Psi_m = 3U_0/P_1$ and $\Delta = \sqrt{4P_2/U_0}$. The solution (43) is a compressive IASWs if $P_1 > 0$ and it is a rarefactive IASWs if $P_1 < 0$. Therefore, the critical condition for transition from the rarefactive IASWs to the compressive IASWs is $P_1 = 0$ in Eq. (36).

When we take $q \to 1$ and $\alpha = 0$, the two-temperature electrons recover to a Maxwellian distribution, and at this time the KdV equation still has the form solution of $sech^2$, while $P_1$ and $P_2$ return to that

$$P_1^M = \frac{l_z}{2}\left(\frac{1-\eta}{1+\eta\nu}+\sigma\gamma\right)^{-\frac{1}{2}}\left\{3-\frac{\sigma\gamma(1+\gamma)(1+\eta\nu)}{-1+\eta}-\frac{(-1+\eta)\left[-1+\eta\nu^2(\mu_c+\beta\mu_h)\right]}{(1+\eta\nu)^2(\mu_c+\beta\mu_h)}\right\},$$

and

$$P_2^M = \frac{l_z}{2(1+\eta\nu)^2}\left(\frac{1-\eta}{1+\eta\nu}+\sigma\gamma\right)^{-\frac{1}{2}}\left[1-\eta+\frac{(1-l_z^2)(1-\eta+\sigma\gamma(1+\eta\nu))^2}{\Omega_{eff}^2}\right]. \tag{44}$$

Therefore, amplitude and width of the solitary wave are, respectively,

$$\Psi_m\big|_{q\to 1,\alpha=0} = 3U_0/P_1^M \quad \text{and} \quad \Delta_{q\to 1,\alpha=0} = \sqrt{4P_2^M/U_0}.$$

3.3 *Numerical analyses*

In order to show the properties of IASWs (43) more clearly, now we make the numerical analyses. For the calculations in the numerical analyses, some appropriate physical parameters of the plasma are chosen as $U_0 = 0.2$, $\mu_{c??} = 0.1$,?? $\mu_h = 0.9$, ??$\eta = 0.52$, $\beta = 0.05$, $\nu = 0.1$, $\gamma = 3$, $\sigma = 1.5$, $l_z = 0.7$, $\omega_{ci} = 0.5$ and $\Omega_0 = 0.01$.

In Figure 2, we showed the critical condition $P_1$ in Eq. (36) as a function of the nonextensive $q$-parameter and the nonthermal $\alpha$-parameter. It is clear that the critical condition can be $P_1 > 0$ or $P_1 < 0$ and therefore the IASWs can be compressive waves or rarefactive waves.

In Figures 3(a)-3(b), we showed the compressive IASWs $\Psi$ in Eq. (43) as a function of $\chi$ for three different $q$-parameters and three different $\alpha$-parameters, respectively, where in the figure 3(a), $\Psi$ as a function of $\chi$ for three different $q$-parameters at a fixed $\alpha = 0.08$, and in the figure 3(b), $\Psi$ as a function of $\chi$ for the three different $\alpha$-parameters at a fixed $q = 0.85$. It is obvious that the amplitude and width of the IASWs decrease as the $q$-parameter increases, and they increase as the $\alpha$-parameter increases.

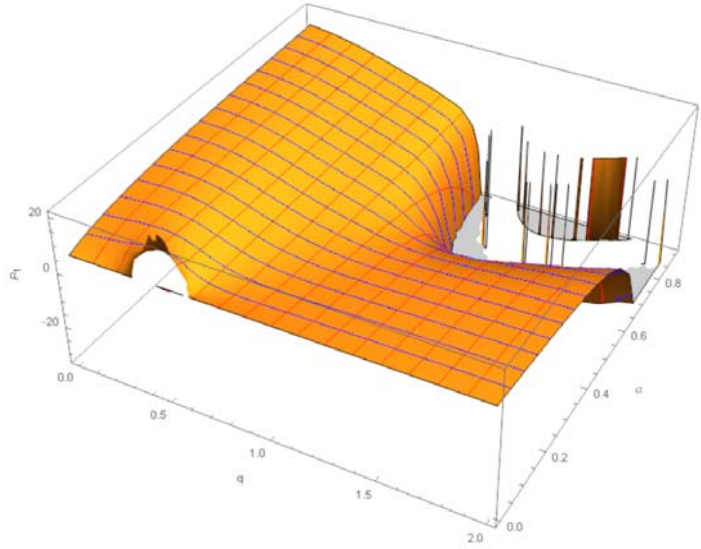

**Figure 2.** Dependence of $P_1$ on *q*-parameter and α-parameter.

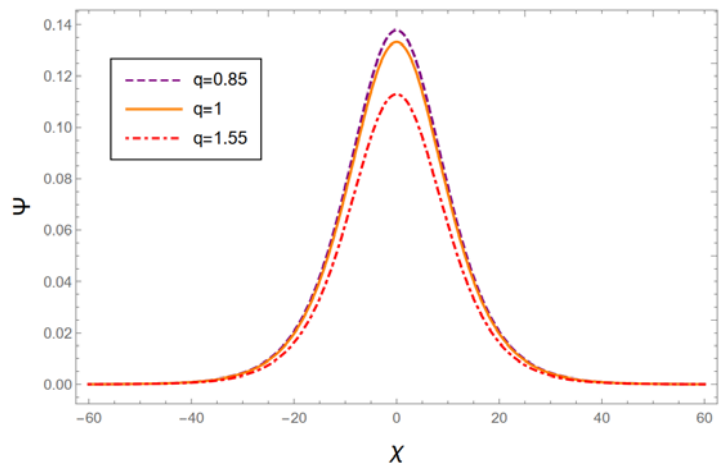

**Figure 3(a).** The compressive IASWs $\Psi$ as a function of $\chi$ for three different *q*-parameters.

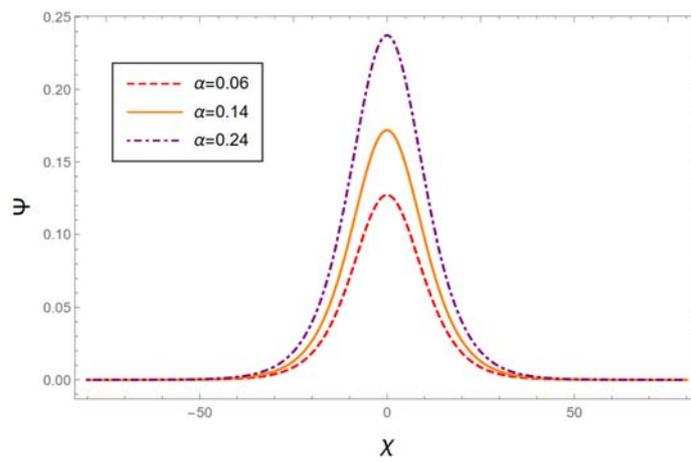

**Figure 3(b).** The compressive IASWs $\Psi$ as a function of $\chi$ for three different α-parameters.

In Figures 4(a)-4(f), when $\Psi$ is as a function of $\chi$, we showed effects of the physical quantities on the compressive IASWs in Eq. (43) at a fixed $q = 1.22$ and a fixed $\alpha = 0.22$, such as the ratio $\sigma$ of the ion temperature to the effective temperature of electrons, the direction cosine $l_z$, the solitary wave velocity $U_0$, the ratio $v$ of the effective temperature to the positron temperature, the density ratio $\eta$ of the positrons to the electrons, and the ion gyrofrequency $\omega_{ci}$, respectively.

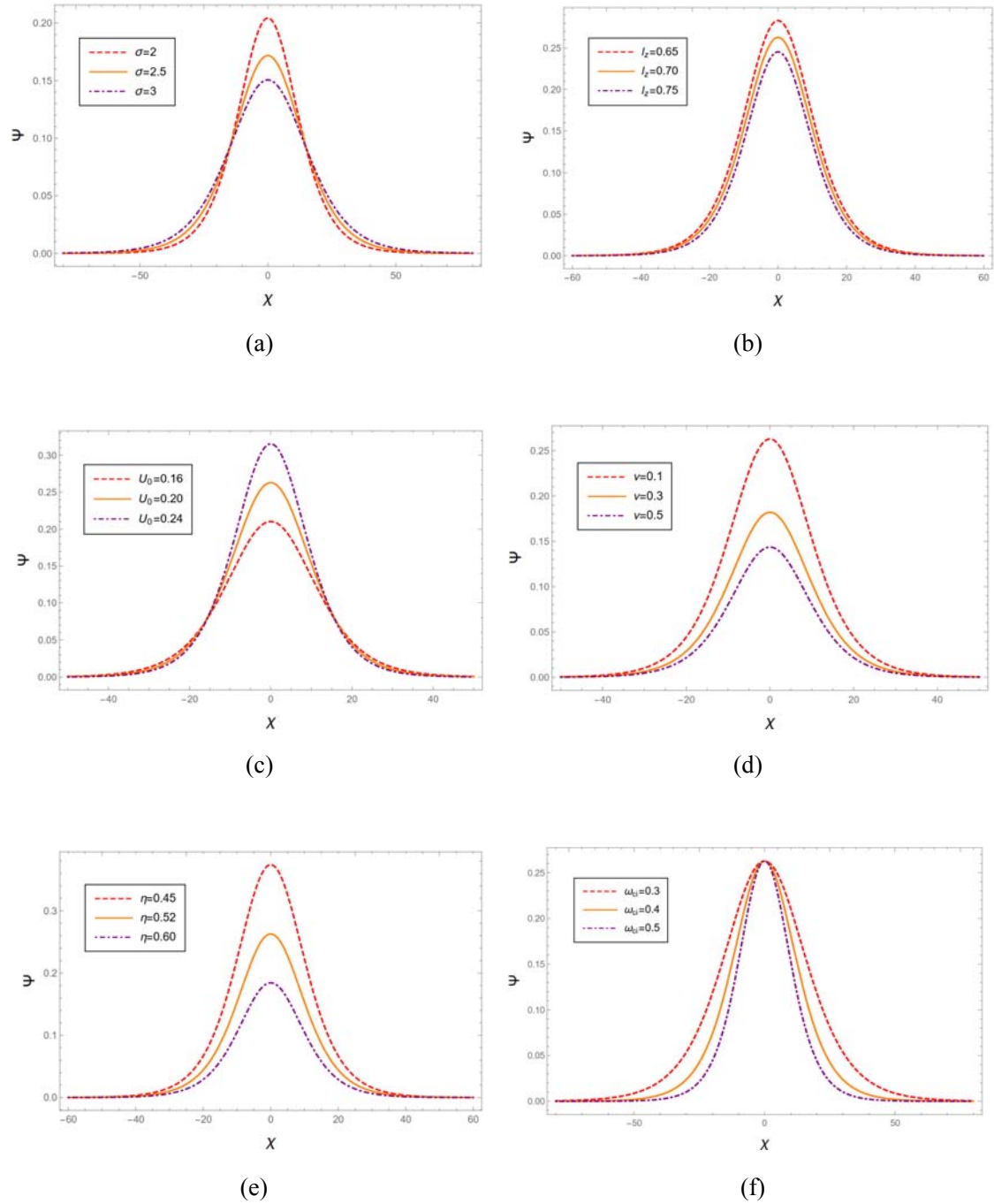

**Figures 4(a)-(f).** The compressive IASWs $\Psi$ as a function of $\chi$ for the different physical quantities, $\sigma$, $l_z$, $U_0$, $v$, $\eta$ and $\omega_{ci}$.

In Figure 4(a), we showed $\Psi$ as a function of $\chi$ for three different ratio $\sigma$, where we find that the amplitude of the compressive IASWs decreases as $\sigma$ increases. In Figure 4(b), we showed $\Psi$ as a function of $\chi$ for three different the direction cosine $l_z$, where we find that the amplitude and width of the compressive IASWs decrease as the direction cosine $l_z$ increases. In Figure 4(c), we showed $\Psi$ as a function of $\chi$ for three different wave speed $U_0$, where we find that the amplitude of the compressive IASWs increases as the wave speed $U_0$ increases. In Figure 4(d) and 4(e), we showed $\Psi$ as a function of $\chi$ for three different ratios $v$ and $\eta$, respectively, where we find that the amplitude and width of the compressive IASWs both decrease as each of $v$ and $\eta$ increases. In Figure 4(f), we showed $\Psi$ as a function of $\chi$ for three different frequency $\omega_{ci}$, where we find that only the width of the compressive IASWs decreases as the frequency $\omega_{ci}$ increases, but the amplitude of the IASWs is unchanged.

## 4. The modified KdV equation and the IASWs

### 4.1 *The equation*

In Section 3, we have derived the KdV equation (35) and its solitary wave solution. In addition to the KdV equation, many other nonlinear partial differential equations have also similar solitary wave solutions. Now, in the plasma model, let us derive the modified Korteweg-de Vries (mKdV) equation, which has small but finite amplitude of IASWs. In the mKdV [51-53], the independent variables are stretched as $(\xi, \tau)$, namely,

$$\xi = \varepsilon(l_x x + l_y y + l_z z - \lambda t), \quad \tau = \varepsilon^3 t, \tag{45}$$

and thus the derivative of coordinates and time become

$$\frac{\partial}{\partial r} = \varepsilon l_r \frac{\partial}{\partial \xi}, \quad \frac{\partial^2}{\partial r^2} = \varepsilon^2 l_r^2 \frac{\partial^2}{\partial \xi^2}, \quad \frac{\partial}{\partial t} = -\lambda \varepsilon \frac{\partial}{\partial \xi} + \varepsilon^3 \frac{\partial}{\partial \tau}. \tag{46}$$

In the same way, the physical quantities $n$, $u_z$, $p$ and $\Psi$ are the same as the power series about the small parameter $\varepsilon$ in Eq. (23), while $u_x$ and $u_y$ are expanded as follows,

$$\begin{cases} u_x = \varepsilon^2 u_x^{(1)} + \varepsilon^3 u_x^{(2)} + \varepsilon^4 u_x^{(3)} + \mathrm{L}, \\ u_y = \varepsilon^2 u_y^{(1)} + \varepsilon^3 u_y^{(2)} + \varepsilon^4 u_y^{(3)} + \mathrm{L}. \end{cases} \tag{47}$$

In the new coordinates $(\xi, \tau)$, the governing Eqs. (16) - (21) are expressed as

$$-\lambda \varepsilon \frac{\partial n_i}{\partial \xi} + \varepsilon^3 \frac{\partial n_i}{\partial \tau} + \varepsilon l_x \frac{\partial (n_i u_x)}{\partial \xi} + \varepsilon l_y \frac{\partial (n_i u_y)}{\partial \xi} + \varepsilon l_z \frac{\partial (n_i u_z)}{\partial \xi} = 0, \tag{48}$$

$$\varepsilon^3 \frac{\partial u_x}{\partial \tau} + \varepsilon \left(-\lambda + l_x u_x + l_y u_y + l_z u_z\right) \frac{\partial u_x}{\partial \xi} + \varepsilon l_x \frac{\partial \Psi}{\partial \xi} - \Omega_{eff} u_y + \varepsilon l_x \frac{\sigma}{n_i} \frac{\partial p}{\partial \xi} = 0, \tag{49}$$

$$\varepsilon^3 \frac{\partial u_y}{\partial \tau} + \varepsilon \left(-\lambda + l_x u_x + l_y u_y + l_z u_z\right) \frac{\partial u_y}{\partial \xi} + \varepsilon l_y \frac{\partial \Psi}{\partial \xi} + \Omega_{eff} u_x + \varepsilon l_y \frac{\sigma}{n_i} \frac{\partial p}{\partial \xi} = 0, \tag{50}$$

$$\varepsilon^3 \frac{\partial u_z}{\partial \tilde{\tau}} + \varepsilon\left(-\tilde{\lambda} + l_x u_x + l_y u_y + l_z u_z\right)\frac{\partial u_z}{\partial \tilde{\xi}} + \varepsilon l_z \frac{\partial \Psi}{\partial \tilde{\xi}} + \varepsilon l_z \frac{\sigma}{n_i}\frac{\partial p}{\partial \tilde{\xi}} = 0, \tag{51}$$

$$\varepsilon^3 \frac{\partial p}{\partial \tilde{\tau}} + \varepsilon\left(-\tilde{\lambda} + u_x l_x + u_y l_y + u_z l_z\right)\frac{\partial p}{\partial \tilde{\xi}} + \varepsilon \gamma p \left(l_x \frac{\partial u_x}{\partial \tilde{\xi}} + l_y \frac{\partial u_y}{\partial \tilde{\xi}} + l_z \frac{\partial u_z}{\partial \tilde{\xi}}\right) = 0, \tag{52}$$

$$\varepsilon^2 \frac{\partial^2 \Psi}{\partial \tilde{\xi}^2} = -\eta + \aleph \Psi + \Re \Psi^2 + \Xi \Psi^3 - (1-\eta)n_i. \tag{53}$$

After substituting Eq. (23) for $n$, $u_z$, $p$ and $\Psi$ and Eq. (47) for $u_x$ and $u_y$ into Eqs. (48)-(53), from the first-terms of the small parameter $\varepsilon$, we can get the same equations as Eqs. (31) and (32). While from the second-order terms of $\varepsilon$, we get the following equations,

$$n_i^{(2)} = \frac{\aleph}{(1-\eta)}\Psi^{(2)} + \frac{\Re}{(1-\eta)}\left[\Psi^{(1)}\right]^2,$$

$$l_z u_z^{(1)} \frac{\partial n_i^{(1)}}{\partial \tilde{\xi}} - \tilde{\lambda}\frac{\partial n_i^{(2)}}{\partial \tilde{\xi}} + l_x \frac{\partial u_x^{(1)}}{\partial \tilde{\xi}} + l_y \frac{\partial u_y^{(1)}}{\partial \tilde{\xi}} + l_z \frac{\partial u_z^{(2)}}{\partial \tilde{\xi}} + l_z n_i^{(1)} \frac{\partial u_z^{(1)}}{\partial \tilde{\xi}} = 0,$$

$$u_y^{(2)} = -\frac{\tilde{\lambda}}{\Omega_{eff}}\frac{\partial u_x^{(1)}}{\partial \tilde{\xi}} + \frac{l_x}{\Omega_{eff}}\frac{\partial \Psi^{(2)}}{\partial \tilde{\xi}} + \frac{l_x}{\Omega_{eff}} n_i^{(1)} \frac{\partial \Psi^{(1)}}{\partial \tilde{\xi}} - n_i^{(1)} u_y^{(1)} + \frac{l_x \sigma}{\Omega_{eff}}\frac{\partial p^{(2)}}{\partial \tilde{\xi}},$$

$$u_x^{(2)} = \frac{\tilde{\lambda}}{\Omega_{eff}}\frac{\partial u_y^{(1)}}{\partial \tilde{\xi}} - \frac{l_y}{\Omega_{eff}}\frac{\partial \Psi^{(2)}}{\partial \tilde{\xi}} - \frac{l_y}{\Omega_{eff}} n_i^{(1)} \frac{\partial \Psi^{(1)}}{\partial \tilde{\xi}} - n_i^{(1)} u_x^{(1)} - \frac{l_y \sigma}{\Omega_{eff}}\frac{\partial p^{(2)}}{\partial \tilde{\xi}},$$

$$-\tilde{\lambda} n_i^{(1)} \frac{\partial u_z^{(1)}}{\partial \tilde{\xi}} + l_z u_z^{(1)} \frac{\partial u_z^{(1)}}{\partial \tilde{\xi}} - \tilde{\lambda}\frac{\partial u_z^{(2)}}{\partial \tilde{\xi}} + l_z \frac{\partial \Psi^{(2)}}{\partial \tilde{\xi}} + l_z n_i^{(1)} \frac{\partial \Psi^{(1)}}{\partial \tilde{\xi}} + l_z \sigma \frac{\partial p^{(2)}}{\partial \tilde{\xi}} = 0,$$

$$l_z u_z^{(1)} \frac{\partial p^{(1)}}{\partial \tilde{\xi}} - \tilde{\lambda}\frac{\partial p^{(2)}}{\partial \tilde{\xi}} + \gamma l_x \frac{\partial u_x^{(1)}}{\partial \tilde{\xi}} + \gamma l_y \frac{\partial u_y^{(1)}}{\partial \tilde{\xi}} + \gamma l_z \frac{\partial u_z^{(2)}}{\partial \tilde{\xi}} + \gamma l_z p^{(1)} \frac{\partial u_z^{(1)}}{\partial \tilde{\xi}} = 0. \tag{54}$$

And from the third-order terms of $\varepsilon$, we get that

$$n_i^{(3)} = \frac{\aleph}{(1-\eta)}\Psi^{(3)} + \frac{2\Re}{(1-\eta)}\Psi^{(1)}\Psi^{(2)} + \frac{\Xi}{(1-\eta)}\left[\Psi^{(1)}\right]^3 - \frac{1}{(1-\eta)}\frac{\partial^2 \Psi^{(1)}}{\partial \tilde{\xi}^2},$$

$$\frac{\partial n_i^{(1)}}{\partial \tilde{\tau}} + \frac{\partial n_i^{(1)}}{\partial \tilde{\xi}}\left(l_x u_x^{(1)} + l_y u_y^{(1)} + l_z u_z^{(2)}\right) + l_z u_z^{(1)} \frac{\partial n_i^{(2)}}{\partial \tilde{\xi}} - \tilde{\lambda}\frac{\partial n_i^{(3)}}{\partial \tilde{\xi}} + l_x \frac{\partial u_x^{(2)}}{\partial \tilde{\xi}} + l_y \frac{\partial u_y^{(2)}}{\partial \tilde{\xi}}$$

$$+ l_z \frac{\partial u_z^{(3)}}{\partial \tilde{\xi}} + l_x n_i^{(1)} \frac{\partial u_x^{(1)}}{\partial \tilde{\xi}} + l_y n_i^{(1)} \frac{\partial u_y^{(1)}}{\partial \tilde{\xi}} + l_z n_i^{(1)} \frac{\partial u_z^{(2)}}{\partial \tilde{\xi}} + l_z n_i^{(2)} \frac{\partial u_z^{(1)}}{\partial \tilde{\xi}} = 0,$$

$$\frac{\partial u_z^{(1)}}{\partial \tilde{\tau}} + l_z n_i^{(1)} u_z^{(1)} \frac{\partial u_z^{(1)}}{\partial \tilde{\xi}} - \tilde{\lambda} n_i^{(2)} \frac{\partial u_z^{(1)}}{\partial \tilde{\xi}} + \frac{\partial u_z^{(1)}}{\partial \tilde{\xi}}\left(l_x u_x^{(1)} + l_y u_y^{(1)} + l_z u_z^{(2)}\right) - \tilde{\lambda} n_i^{(1)} \frac{\partial u_z^{(2)}}{\partial \tilde{\xi}}$$

$$+ l_z u_z^{(1)} \frac{\partial u_z^{(2)}}{\partial \tilde{\xi}} - \tilde{\lambda}\frac{\partial u_z^{(3)}}{\partial \tilde{\xi}} + l_z \frac{\partial \Psi^{(3)}}{\partial \tilde{\xi}} + l_z n_i^{(1)} \frac{\partial \Psi^{(2)}}{\partial \tilde{\xi}} + l_z n_i^{(2)} \frac{\partial \Psi^{(1)}}{\partial \tilde{\xi}} + l_z \sigma \frac{\partial p^{(3)}}{\partial \tilde{\xi}} = 0,$$

$$\frac{\partial p^{(1)}}{\partial \tilde{\tau}} + \frac{\partial p^{(1)}}{\partial \tilde{\xi}}\left(l_x u_x^{(1)} + l_y u_y^{(1)} + l_z u_z^{(2)}\right) + l_z u_z^{(1)} \frac{\partial p^{(2)}}{\partial \tilde{\xi}} - \tilde{\lambda}\frac{\partial p^{(3)}}{\partial \tilde{\xi}} + \gamma l_x \frac{\partial u_x^{(2)}}{\partial \tilde{\xi}} + \gamma l_y \frac{\partial u_y^{(2)}}{\partial \tilde{\xi}}$$

$$+ \gamma l_z \frac{\partial u_z^{(3)}}{\partial \tilde{\xi}} + l_x \gamma p^{(1)} \frac{\partial u_x^{(1)}}{\partial \tilde{\xi}} + l_y \gamma p^{(1)} \frac{\partial u_y^{(1)}}{\partial \tilde{\xi}} + l_z \gamma p^{(1)} \frac{\partial u_z^{(2)}}{\partial \tilde{\xi}} + \gamma l_z p^{(2)} \frac{\partial u_z^{(1)}}{\partial \tilde{\xi}} = 0. \tag{55}$$

Now we express the following high-order quantities as a function of $\Psi^{(1)}$, namely,

$$u_x^{(2)} = \frac{l_x}{\Omega_{eff}^2}\left(\frac{\tilde{\lambda}^2}{\tilde{\lambda}^2-\sigma\gamma l_z^2}\right)\frac{\partial^2\Psi^{(1)}}{\partial\xi^2} - \frac{l_y}{\Omega_{eff}}\left(\frac{\tilde{\lambda}}{\tilde{\lambda}^2-\sigma\gamma l_z^2}\right)\frac{\partial\Psi^{(2)}}{\partial\xi} - \frac{l_y}{\Omega_{eff}}\frac{\tilde{\lambda}\sigma\gamma(1+\gamma)l_z^4}{\left(\tilde{\lambda}^2-\sigma\gamma l_z^2\right)^3}\Psi^{(1)}\frac{\partial\Psi^{(1)}}{\partial\xi},$$

$$u_y^{(2)} = \frac{l_y}{\Omega_{eff}^2}\left(\frac{\tilde{\lambda}^2}{\tilde{\lambda}^2-\sigma\gamma l_z^2}\right)\frac{\partial^2\Psi^{(1)}}{\partial\xi^2} + \frac{l_x}{\Omega_{eff}}\left(\frac{\tilde{\lambda}}{\tilde{\lambda}^2-\sigma\gamma l_z^2}\right)\frac{\partial\Psi^{(2)}}{\partial\xi} + \frac{l_x}{\Omega_{eff}}\frac{\tilde{\lambda}\sigma\gamma l_z^4(1+\gamma)}{\left(\tilde{\lambda}^2-\sigma\gamma l_z^2\right)^3}\Psi^{(1)}\frac{\partial\Psi^{(1)}}{\partial\xi},$$

$$u_z^{(2)} = \frac{\tilde{\lambda}l_z}{\tilde{\lambda}^2-\sigma\gamma l_z^2}\Psi^{(2)} + \frac{\tilde{\lambda}l_z^3}{\left(\tilde{\lambda}^2-\sigma\gamma l_z^2\right)^2}\frac{1}{2}\left[\Psi^{(1)}\right]^2 + \frac{\sigma\gamma\tilde{\lambda}l_z^5(1+\gamma)}{\left(\tilde{\lambda}^2-\sigma\gamma l_z^2\right)^3}\frac{1}{2}\left[\Psi^{(1)}\right]^2,$$

$$\frac{\partial u_z^{(3)}}{\partial\xi} = \frac{l_z\left(\tilde{\lambda}^2+\sigma\gamma l_z^2\right)}{\left(\tilde{\lambda}^2-\sigma\gamma l_z^2\right)^2}\frac{\partial\Psi^{(1)}}{\partial\tau} + \frac{\tilde{\lambda}l_z}{\tilde{\lambda}^2-\sigma\gamma l_z^2}\frac{\partial\Psi^{(3)}}{\partial\xi} + \frac{\tilde{\lambda}l_z\sigma}{\left(\tilde{\lambda}^2-\sigma\gamma l_z^2\right)^2}\frac{\gamma}{\Omega_{eff}^2}\left(1-l_z^2\right)\frac{\partial^3\Psi^{(1)}}{\partial\xi^3}$$

$$+\frac{\tilde{\lambda}l_z^3}{\left(\tilde{\lambda}^2-\sigma\gamma l_z^2\right)^2}\left(1+\frac{\sigma\gamma l_z^2(1+\gamma)}{\tilde{\lambda}^2-\sigma\gamma l_z^2}\right)\left(\Psi^{(1)}\frac{\partial\Psi^{(2)}}{\partial\xi}+\Psi^{(2)}\frac{\partial\Psi^{(1)}}{\partial\xi}\right) + D\left[\Psi^{(1)}\right]^2\frac{\partial\Psi^{(1)}}{\partial\xi}. \quad (56)$$

where we denote

$$D = \frac{\Re\left(\tilde{\lambda}^3 l_z - \tilde{\lambda}\sigma\gamma l_z^3\right)}{(1-\eta)\left(\tilde{\lambda}^2-\sigma\gamma l_z^2\right)^2} + \frac{3\tilde{\lambda}^3 l_z^5}{2\left(\tilde{\lambda}^2-\sigma\gamma l_z^2\right)^4} + \frac{\sigma\gamma\tilde{\lambda}^3 l_z^7(1+\gamma)}{2\left(\tilde{\lambda}^2-\sigma\gamma l_z^2\right)^5}$$

$$+\frac{\tilde{\lambda}\sigma\gamma l_z^7\left(\frac{5+\gamma^2}{2}+3\gamma\right)}{\left(\tilde{\lambda}^2-\sigma\gamma l_z^2\right)^4} + \frac{3\tilde{\lambda}\sigma^2\gamma^2 l_z^9(1+\gamma)^2}{2\left(\tilde{\lambda}^2-\sigma\gamma l_z^2\right)^5}.$$

From Eqs. (54) and (55), we eliminate all the higher-order perturbation terms, and then from the above equations we finally derive the mKdV equation for IAWs,

$$\frac{\partial\Psi^{(1)}}{\partial\tau} + P_3\left(\Psi^{(1)}\right)^2\frac{\partial\Psi^{(1)}}{\partial\xi} + P_2\frac{\partial^3\Psi^{(1)}}{\partial\xi^3} = 0, \quad (57)$$

where the nonlinear coefficient $P_3$ is given by

$$P_3 = \frac{3\Xi\left(\tilde{\lambda}^2-\sigma\gamma l_z^2\right)^2}{2\tilde{\lambda}l_z^3(-1+\eta)} + \frac{15\tilde{\lambda}l_z^4 + \tilde{\lambda}\sigma\gamma l_z^6\left[-18+\gamma(13+\gamma)\right]+\sigma^2\gamma^2 l_z^8(-2+\gamma)(-3+2\gamma)}{4\tilde{\lambda}\left(\tilde{\lambda}^2-\sigma\gamma l_z^2\right)^3}, \quad (58)$$

and the dispersion coefficient $P_2$ is the same as that in Eq. (37).

4.2 *The IASWs*

For the IASWs solution, we still use the traveling wave method to solve Eq. (57). Let $\Psi^{(1)} = \Psi$ and $\chi = \xi - U_0\tau$, the mKdV equation (57) becomes

$$-U_0\frac{d\Psi}{d\chi} + P_3\Psi^2\frac{d\Psi}{d\chi} + P_2\frac{d^3\Psi}{d\chi^3} = 0, \quad (59)$$

Integrating Eq. (59) for $\chi$, and then multiplying by $d\Psi/d\chi$ and integrating it again, we yield

$$-\frac{1}{2}U_0\Psi^2 + \frac{1}{12}P_3\Psi^4 + \frac{1}{2}P_2\left(\frac{d\Psi}{d\chi}\right)^2 = c_4\Psi + c_5, \quad (60)$$

where $c_4$ and $c_5$ are the integration constants. According to the boundary condition (40), we can determine the constants, $c_4 = c_5 = 0$. Thus, Eq. (60) is written as

$$\left(\frac{d\Psi}{d\chi}\right)^2 = \frac{P_3}{6P_2}\Psi^2\left(\frac{6U_0}{P_3} - \Psi^2\right). \tag{61}$$

It can be seen that $\Psi$ is a real function only if $\Psi^2 < 6U_0/P_3$. By separating the variables and then integrating them, we obtain the soliton solution,

$$\Psi = \pm\sqrt{\frac{6U_0}{P_3}}\text{sech}\left(\chi\sqrt{\frac{U_0}{P_2}}\right), \tag{62}$$

where the positive and negative signs are respectively associated to the compressive and rarefactive IASWs.

When we take $q \to 1$ and $\alpha = 0$, i.e., the electron velocity distribution returns to a Maxwellian distribution, the mKdV equation has still the *sech* form solution, but the coefficient $P_2$ becomes $P_2^M$ and the coefficient $P_3$ becomes

$$P_3^M = \frac{l_z}{4(\eta-1)^3(1+\eta\nu)^2}\left(\frac{1-\eta}{1+\eta\nu}+\sigma\gamma\right)^{-\frac{1}{2}}\left\{\sigma^2\gamma^2(2-\gamma)(2\gamma-3)(1+\eta\nu)^5\right.$$
$$-\sigma\gamma(\gamma^2+13\gamma-18)(1+\eta\nu)^4\left[1-\eta+\sigma\gamma(1+\eta\nu)\right]-15(1+\eta\nu)^3\left[1-\eta+\sigma\gamma(1+\eta\nu)\right]^2 \tag{63}$$
$$\left.+\frac{(1-\eta)^4\left[1+\eta\nu^3(\mu_c+\beta\mu_h)^2\right]}{(\mu_c+\beta\mu_h)^2}\right\}.$$

Correspondingly, the amplitude and the width are given, respectively, by

$$\Psi_m\big|_{q\to 1,\alpha=0} = \sqrt{6U_0/P_3^M} \quad \text{and} \quad \Delta_{q\to 1,\alpha=0} = \sqrt{P_2^M/U_0}.$$

*4.3 The numerical analyses*

Based on the IASWs solution (62), we can make the numerical analyses of the properties of the IASWs when they are compressive. In the numerical analyses, the values of the basic plasma physical quantities for the numerical calculations are chosen the same as those in Section 3.

In Figures 5(a) and 5(b), we showed the roles of the $q$-parameter and the $\alpha$-parameter in the IASWs, where the figure 5(a) showed $\Psi$ as a function of $\chi$ for three different values of $q$ at a fixed $\alpha = 0.06$, the figure 5(b) showed $\Psi$ as a function of $\chi$ for three different values of $\alpha$ at a fixed $q = 1.22$. We find that the amplitude and width of the IASWs decrease with the increase of the parameter $q$, but increase with the increase of the parameter $\alpha$.

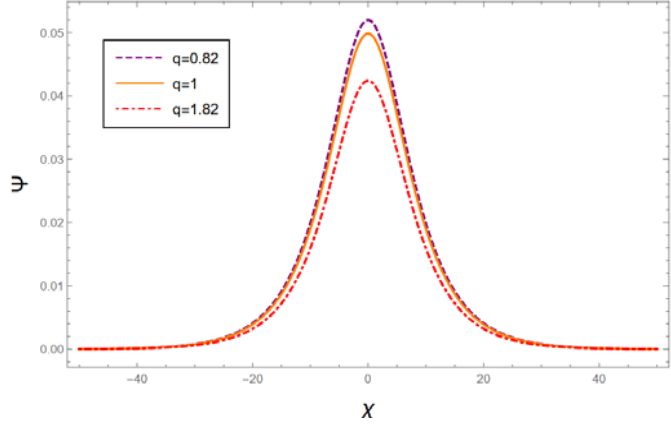

**Figure 5(a)**. Ψ as a function of χ for the three different *q*-parameters

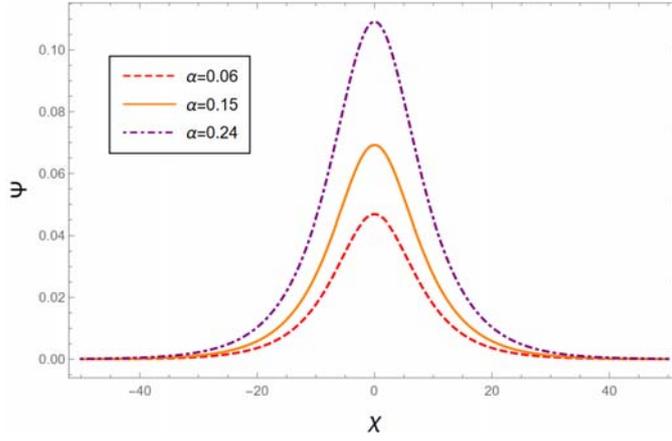

**Figure 5(b).** Ψ as a function of χ for the three different α-parameters.

In Figures 6(a)-6(f), we showed the effects of the plasma physical quantities on the IASWs, such as the ratio $\sigma$ of the ion temperature to the effective temperature of electrons, the direction cosine $l_z$, the solitary wave velocity $U_0$, the ratio $\nu$ of the effective temperature to the positron temperature, the density ratio $\eta$ of the positrons to the electrons and the ion gyrofrequency $\omega_{ci}$, where the *q*- and α-parameter are fixed at $q = 1.22$ and $\alpha = 0.15$.

The Figure 6(a) showed Ψ as a function of χ for three different values of $\sigma$, where we find that as $\sigma$ increases, the amplitude of the IASWs decreases but the width increases. The Figure 6(b) showed Ψ as a function of χ for three different values of $l_z$, where we find that the amplitude and the width of the IASWs decrease with the increase of $l_z$. The Figure 6(c) showed Ψ as a function of χ for three different values of $U_0$, where we find that the amplitude of the IASWs increases as the wave speed $U_0$ increases. The Figures 6(d) and 6(e) showed Ψ as a function of χ for three different values of $\nu$ and of $\eta$, respectively, where we find that the amplitude and width of the IASWs decrease as the two quantities $\nu$ and $\eta$ increase. The Figure 6(f) showed Ψ as a function of $\chi$ for three different values of $\omega_{ci}$, where we find that as $\omega_{ci}$ increases, the width of the IASWs decreases but the amplitude is unchanged.

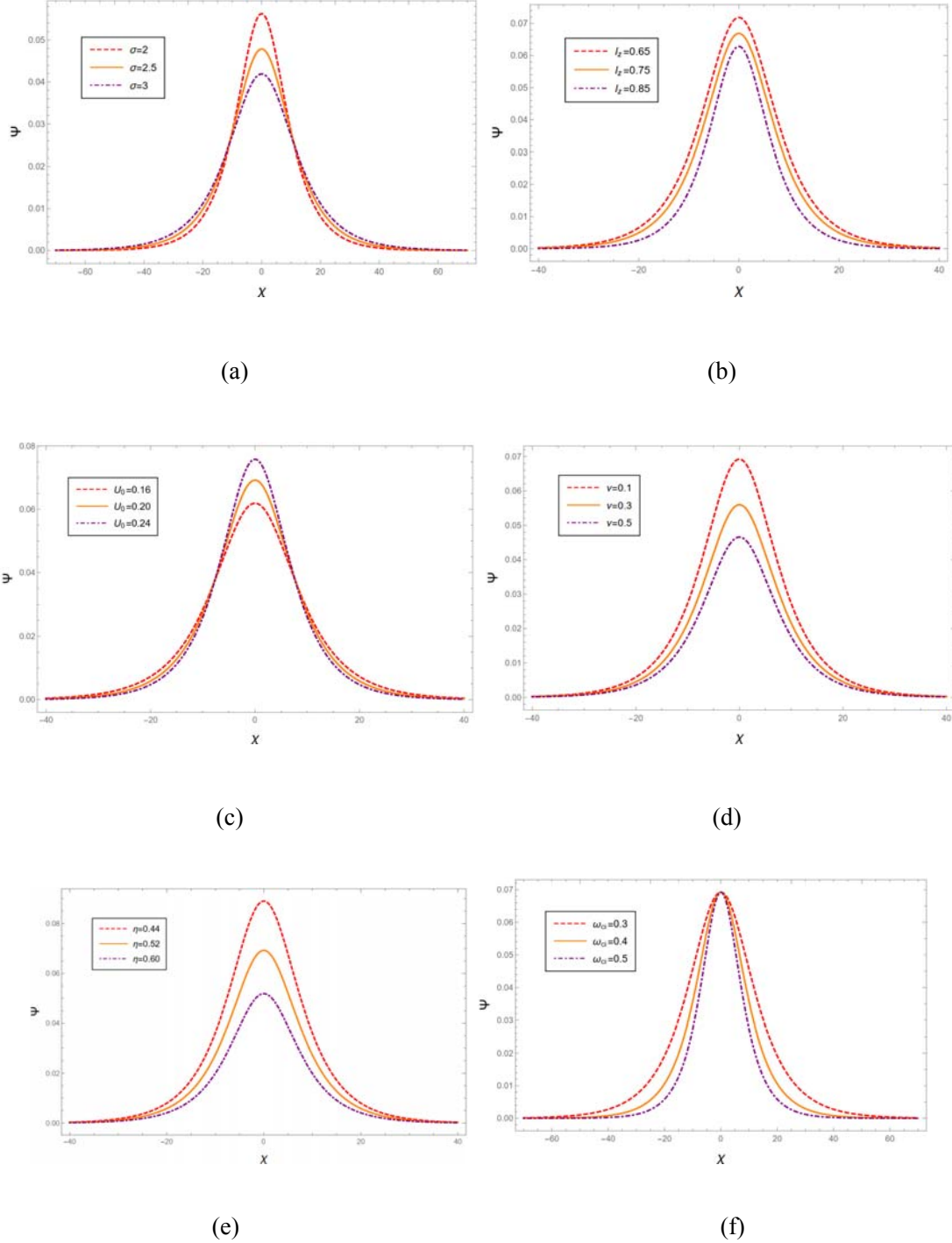

**Figures 6(a)-(f).** The compressive IASWs $\Psi$ as a function of $\chi$ for the three different values of the plasma physical quantities $\sigma$, $l_z$, $U_0$, $v$, $\eta$ and $\omega_{ci}$.

## 4. Conclusions

In conclusions, we have studied small amplitude IASWs in the four-component magneto-rotating plasma with cold fluid ions, hot positrons and two-temperature electrons (cold and hot electrons), where the two-temperature electrons obey the modified Cairns-Tsallis distribution, a non-Maxwell distribution. We have derived the KdV equation and the mKdV equation for the multi-component plasma model by employing the RPM, which are given by Eq. (35) and Eq. (57), respectively. From the KdV equation and the mKdV equation, we have used the traveling wave method to derive their small amplitude IASWs solutions described by Eq. (43) and Eq. (62), respectively. We find that the properties of the small amplitude IASWs depend significantly on the $q$-parameter and the $\alpha$-parameter in the modified Cairns-Tsallis distribution, and thus differ from the cases in the plasma with a Maxwellian distribution. The IASWs properties depend also on the plasma physical quantities, such as the ratio $\sigma$ of the ions temperature and the effective temperature of electrons, the direction cosine $l_z$, the solitary wave velocity $U_0$, the ratio $\nu$ of the effective temperature to the positron temperature, the ratio $\eta$ of the positron density to the electron density and the ion gyrofrequency $\omega_{ci}$ etc.

In addition, we have showed that the small amplitude IASWs are compressive if the nonlinear coefficient $P_1$ in the KdV Eq. (35) is positive, $P_1 > 0$, or the IASWs are rarefactive if the coefficient is $P_1 < 0$. The nonlinear coefficient $P_1$ is given by Eq. (36). The numerical analysis of the coefficient $P_1$ is made as a function of the parameters $q$ and $\alpha$ in the modified Cairns-Tsallis distribution, which is shown in Figure 2.

Further, we have made numerical analyses of the small amplitude IASWs solutions $\Psi$ in Eq. (43) and Eq. (62), which are derived from the KdV equation and the mKdV equation, respectively. In Figures 3-6, we showed the properties of the IASWs solutions $\Psi$ as a function of the generalized coordinate $\chi$ for various physical cases in the plasma. We illustrated the significant effects of the modified Cairns-Tsallis distribution on the IASWs. We also illustrated the roles of the plasma physical quantities, such as $\sigma$, $l_z$, $U_0$, $\nu$, $\eta$ and $\omega_{ci}$, in the IASWs. All the results are important for us to understand the IASWs in the multi-component astrophysical and space plasmas with the non-Maxwellian distributions.

**Acknowledgements**

This work was supported by the National Natural Science Foundation of China under Grant No. 11775156. Guo Ran was supported by the National Natural Science Foundation of China under Grant No. 12105361.

**Data availability statement**

All data that support the findings of this study are included within the article (and any supplementary files).

**Conflicts of interest**

This work was published without conflict of interest.